\documentstyle[preprint,aps,epsf]{revtex}
\newcommand{\deltabar}{\,\,{\bar{}\hspace{1pt}\! \!\delta }}
\newcommand{\dbar}{\,\,{\bar{}\hspace{1pt} \!\!\!d }}
\newcommand{\varepsilonb}{{\bf \varepsilon}}
\begin{document}
\sloppy

\newcommand{\lfrac}[2]{#1/#2}
\renewcommand{\baselinestretch}{1.3}       %

\title{\vspace{-5.5cm} ~ \hfill {\normalsize FUB-HEP/94-5}\\[4.5cm]
Motion of a Rigid Body \\
in Body-Fixed Coordinate System ---
\\an Example
for Autoparrallel Trajectories
in Spaces with Torsion
   }
\author{P.~Fiziev\thanks{Permanent address:
 Department of Theoretical Physics,
 Faculty of Physics, University of Sofia,
 Bull.~5 James Boucher, Sofia~1126,
 Bulgaria.
 Work supported by the Commission of the
 European Communities for Cooperation in Science
 and Technology, Contract~No. ERB3510PL920264
 }
 and H.~Kleinert}
\address{Institut f\"{u}r Theoretische Physik,\\
          Freie Universit\"{a}t Berlin\\
          Arnimallee 14, D - 14195 Berlin}
\date{\today}
\maketitle
\begin{abstract}
We use a recently developed
action principle in spaces with curvature and torsion to
derive
 the Euler equations of motion for a rigid body
   within the body-fixed coordinate system.
This serves as an example that the particle trajectories
in a space with curvature and torsion follow the straightest paths
 (autoparallels),
not the shortest paths (geodesics), as commonly believed.
\end{abstract}
\newpage
\noindent
{\bf 1)~Introduction}\\
Since their discovery more than two centuries ago,
the Euler equations
\begin{equation}
\dot{\bf L} + {\bf  \Omega } \times {\bf L} = {\bf M}
\label{1}\end{equation}
have played a key role
 in understanding the rotations
 of a rigid body around a fixed point O.
 (see for example \cite{1.} -- \cite{4.} and the references therein).
The vectors
   refer to the body-fixed system, ${\bf L}$ being the
 angular momentum,
${\bf  \Omega }$
 the instantaneous angular velocity, and
${\bf M}$  the moments of the external forces. The dot
denotes  differentiation with respect to the time $t$.
The unit basis vectors
$\varepsilonb _i (i = 1,2,3)$
 in B may be assumed to point along the eigenvectors of
 the inertia tensor with respect to O.
Then the angular moment ${\bf L}$ has the components
$
L_i = I_i  \Omega ^i
$ (no sum over $i$).

To describe also the translational
 motion of the rigid body one usually chooses the point O
to coincide with the center of mass
 moving through space.
The motion satisfies the
  additional
equations:
\begin{equation}
 \dot{\bf P} + {\bf  \Omega }
      \times {\bf P} = {\bf F}
\label{2}\end{equation}
where ${\bf P} = M{\bf V}$
is the linear momentum, $M$ the body's
total mass, ${\bf V}$ the velocity
of the center of mass,
and ${\bf F}$ the driving force.
Presently, the equations  (\ref{1}) and (\ref{2})  play
an important role in missile dynamics analysis \cite{4.}.

Usually, the two sets of
 equations are derived as the  Newton equations
 of motion in the {\em  stationary\/}  reference system $S$:
\begin{eqnarray}
& &\dot{\bf l}  =  {\bf m},\label{3b}\\
& &       \dot{\bf p}  = {\bf f}.  \label{3a}
\end{eqnarray}
Following general conventions,
we  distinguish vectors and tensors in $S$
and $B$ by using for the small letters with Greek
indexes and the  capital letters
 with Latin indexes, respectively.

An transformation %
\begin{eqnarray}
 \Omega ^i& = &e^i{}_ \mu  \omega ^ \mu ,\label{4}\\
 V^i & = & \varepsilonb^i {}_ \mu v^ \mu  \label{4a},
\end{eqnarray}
carries the of the angular and linear velocities
$ \omega ^ \mu $ and
$v^ \mu $ to the moving body-fixed system B
 (see \cite{1.} -- \cite{6.}).

The $3 \times 3$ matrix elements  $e^i{}_ \mu $
and $\varepsilon^i{}_ \mu $
  depend on the coordinates
$q^ \mu $ in the system $S$. They satisfy
 $ \partial_{[ \nu } e^i{}_{ \mu ]}  \neq 0$,
$\partial _{[\nu} \varepsilon^i{}_{ \mu ]} \neq 0$,
making the transformation (\ref{4})
and (\ref{4a})
{\em anholonomic
coordinate
transformations\/}  (see
\cite{1.}--\cite{arnold} for the general
theory of the dynamics in
anholonomic coordinates is presented).

Under the transformation (\ref{4}), the equations
(\ref{3b}) go over into
(1) and the additional term $ ({\bf  \Omega } \times {\bf L})^k =
 \epsilon^k{}_{ij}  \Omega ^i L^j $
arises as the moment of the gyroscopic force.
The additional term
$({\bf  \Omega } \times {\bf P} )^k =  \epsilon^k{} _{ij}
 \Omega ^i P^j$ in  equation  (\ref{2})  arise
similarly.
Both terms are a consequence of the anholonomy
of the transformation (\ref{4}), (\ref{4a}).

Within the stationary
system $S$, Hamilton's action principle serves to
derive the equations  of motion (\ref{3b}) and
 (\ref{3a}).
 If one transforms the classical action to the system $B$,
 however, the description involves nonholonomic coordinates and
a naive application of Hamilton's principle
 produces
 wrong equations of motion lacking the additional
gyroscopic moments.
In 1901, Poincar\'e showed \cite{poinc} how to vary an action expressed in terms
 of nonholonomic coordinates.
Following his treatment, one certainly recovers the gyroscopic forces.
His treatment is reviewed in the Textbook \cite{arnold}.

The purpose of this note is to point out that  Poincar\'e's
treatment may be viewed geometrically
as an application of a recently proposed action principle in spaces with
 curvature
and torsion to the spinning top within the body-fixed reference system.
The motivation for such a consideration derives from the fact that,
in the literature on gravity with curvature and torsion \cite{hehl} (for the
 geometry of
such spaces see
\cite{2.}), there
is a widespread belief that in spaces with
torsion, spinless particles move
on shortest paths. However, it was discovered
in Ref. \cite{8.}  (when solving the path integral of the hydrogen atom)
that the correct trajectories are the straightest paths
in a given geometry.
In Ref. \cite{9.},
a classical action principle was found
to comply with this physical fact.

The key to the new action principle
is the observation, that
a space ${\rm M}_q$ with nonzero Riemann
 curvature and torsion may be mapped locally
into a euclidean space ${\rm M}_x$
by an anholonomic transformation
$\dot x^i = e^i{}_\mu \dot q^\mu$ \cite{7.}, \cite{8.}.
For the particle with mass $m$ in ${\rm M}_x$, the equations
of motion $\ddot x^i = 0$ yield straight-line
trajectories. Under an anholonomic transformation, these
lines go over into autoparallels satisfying the equation
$\ddot q^\mu +  \Gamma _{ \alpha  \beta }{}^ \mu
 \dot q^ \alpha \dot q^ \beta  = 0$ with  $\Gamma _{  \alpha \beta }
{}^ \mu  = e_i{}^ \mu  \partial _ \alpha e^i{}_ \beta $
being the Cartan connection.

The orbits in the euclidean space ${\rm M}_x$ may be derived
from Hamilton's principle $ \delta A_x = 0$ applied
to the classical action $A_x = \int^{t_2}_{t_1} dt  \frac{1}{2}m
\dot x^i \dot x^i$.
 Under the anholonomic transformation
$\dot x^i = e^i{}_ \mu \dot q^\mu $, this action goes into
the curvilinear form $ A_q = \int^{t_2}_{t_1} dt \frac{ 1}{2}m
g_{ \mu  \nu } \dot q^ \mu \dot q^ \nu $.
In the space ${\rm M}_q$, a naive application of Hamilton's principle
$ \delta A_q = 0$ produces  wrong equations of motion
in the space $\rm{ M}_q$.
One finds $\ddot q^ \mu  + \bar  \Gamma _{ \alpha \beta }{}
^\mu \dot q^ \alpha  \dot q^ \beta  = 0$
 which are the
 equations for the geodesics rather than the autoparallels;
they lack
 torsion force $ 2m S_ \alpha {}^ \mu {}_ \beta  \dot q^ \alpha
  \dot q^ \beta $.

The problem of describing the dynamics of a rigid body within the
body-fixed frame may be formulated is somewhat analogous.
Here the parameter space of Euler angles has a Riemann curvature.
By going to the body-fixed frame and describing the system in terms of
non-holonomic coordinates,
the space becomes affine-flat but possesses torsion.
The affine connection
in the space of anholonomic coordinates.
is obtained from the spatial derivatives
of
the transformation matrices in
(5) and (6).
The associated
Cartan curvature tensor $R_{  \mu  \nu  \lambda  }{}^  \kappa
\equiv e_i{}^  \kappa (\partial _ \mu \partial _\nu - \partial_ \nu  \partial _
 \mu )
e^i_ \lambda
 $
vanishes, making the connection
affine-flat.
   The antisymmetric part of the connection
 $S_{ \alpha  \beta }{}^ \gamma
=  \frac{1}{2}(
\Gamma _{ \alpha  \beta }{}^ \gamma
-\Gamma _{ \beta \alpha  }{}^ \gamma )
$, which is a tensor, is
nonzero giving
rise to a nonvanishing Riemann curvature tensor
 $\bar{R}_{ \mu  \nu  \lambda  }{}^ \kappa   \neq 0$.
The latter is
formed from the Levi-Cevita connection, also
called Christoffel symbol
$\bar \Gamma _{  \mu  \nu  \lambda  }= \frac{1}{2}
\left(\partial _  \mu  g_{ \nu  \lambda  }+
 \partial _  \nu  g_{ \mu  \lambda  }- \partial _  \lambda
  g_{ \mu  \nu  } \right)$, where $g_{ \mu  \nu  } =
  e^i{}_  \mu   e^i{}_  \nu  $ is the Riemann metric
 in the space of anholonomic coordinates.

We shall apply the variational principle of Ref. \cite{9.}
and derive,
within the body-fixed reference system B,
both the Euler equations (1) and  equations
for the translational motion
(2).
\\[5mm]
{\bf 2)~The ${\rm SO}(3)$ geometry in
 a rotating
  body-fixed system produced by an anholonomic
transformation from the rest system}\\
Consider first a rigid body rotating
around  a fixed point.
Within  the body-fixed system B,
we
 introduce anholonomic coordinates
 $ \Phi  ^i$
corresponding to the transformation  (\ref{4}).
 They are the components of the body's
rotation vector in the axis-angle parametrization.
 The anholonomic transformation  (\ref{4})
defines their infinitesimal increments $ \dbar \Phi ^i =
  \Omega ^i dt$.
For a precise specification, let us go to the system
$S$ where
  the standard Euler angles $ \alpha =  \varphi ^3,  \beta
=  \varphi ^2,~
 \gamma  =  \varphi ^1$ [10]
parametrize (holonomically)
 the body's configuration space ${\rm M}^{(3 )}= {\rm SO}(3)$. The
components of the
 angular velocity in the system B
are $  \Omega ^1 = s_1 \dot\varphi^2 - c_1 s_2 \dot\varphi^3,~
 \Omega ^2 = c_1 \dot\varphi^2 + s_1 s_2 \dot\varphi^3 , \Omega ^3
=   \dot\varphi^1 + c_2 \dot\varphi^3 $
	 (with the short notation $c_ \mu  = \cos  \varphi ^ \mu ,~
s_ \mu  = \sin  \varphi ^ \mu $).
 The basic relation
\begin{equation}
\dbar  \Phi ^i = e^i{}_ \mu  (  \varphi ) d  \varphi ^ \mu
\label{5}\end{equation}
%
has coefficients $e^i{}_ \mu $ which form the matrix
\begin{equation}
 e= \left(
\begin{array}{ccr}
 0 & s_1 &  -c_1 s_2\\
  0 & c_1 &   s_1 s_2 \\
  1     & 0   & c_2
\end{array}
\right).
\label{7}\end{equation}
The symbol $\dbar$
 stands for increments do not
belong to an integrable function; they do not
 satisfy the Schwarz integrability condition
which would read,
  with the coefficients of (7),
$\partial _ \mu e^i {}_ \nu  - \partial _ \nu e^i{}_ \mu
  = 0$.

The metric  $ g = \left( g_{ \alpha  \beta } \right) $
and its inverse  $ g^{-1} = \left( g^{ \alpha  \beta }\right) $
have the matrices
\begin{equation}
 g = \left(
\begin{array}{ccc}
 1 & 0 & c_2\\
 0 & 1 & 0 \\
  c_2 & 0 & 1
\end{array}\right), \,\,\,
g^{-1}=\left(
\begin{array}{ccc}
1 & 0 & -c_2\\
0 & s_2{}^2 & 0 \\
-c_2 & 0 & 1
\end{array}
\right),
\label{x8}\end{equation}
where $|g| = \det  g = s_2{}^2;~~~ g^{-1} = |g|^{-1}$
is the determinant of $g$.
From $g$ we calculate the  Christoffel symbol:
$$
\bar  \Gamma _{ \mu  \nu  \lambda } =
 - \frac{  \sqrt{|g|} }{2} \left( \delta _{1 \mu }
    \delta _{2 \nu } \delta _{3 \lambda } +
    \delta _{1 \nu }  \delta _{2 \mu }  \delta _{3 \lambda }
    +  \delta _{1 \lambda }   \delta _{2 \mu }
     \delta _{3 \nu } +   \delta _{1 \lambda }
       \delta _{2 \nu }  \delta _{3 \mu }
      -  \delta _{1 \mu }  \delta _{2 \lambda }
           \delta _{3 \nu  } -  \delta _{1 \nu }  \delta _{2 \lambda }
           \delta _{3 \mu }\right).
$$
The associated Riemannian Ricci tensor is
$\bar R_{ \alpha  \beta } = \frac{1}{2} g_{ \alpha  \beta }$
with a constant nonzero scalar Riemannian curvature
 $\bar R = \lfrac{3}{2}$.
The only nonzero components of the affine-flat Cartan connection are
\begin{equation}
   \Gamma _{12}{}^1 = ~~ \Gamma _{23}{}^3 =  \frac{c_2}{s_2},
     ~~~\Gamma _{12}{}^3  = ~~~ \Gamma _{23}{}^1 = -\frac{1}{s_2},
     ~~~  \Gamma _{13}{}^{2} = s_2.
\label{9}\end{equation}
From these we find
the antisymmetric part is the torsion tensor $S_{ \alpha  \beta  \gamma }
   = - \frac{1}{2}  \sqrt{|g|}  \epsilon _{ \alpha  \beta  \gamma }$.
 A transforming to
 the local basis $\varepsilonb^i$ yields
the so-called object of anholonomy
\cite{1.}--\cite{3.}
\begin{equation}
 2S_{ijk} = 2e_i^ \alpha  e^ \beta _j e_k^ \mu  S_{ \alpha  \beta  \gamma }
\label{10a}\end{equation}
whose explicit form is
\begin{equation}
2 S_{ijk} = - \epsilon _{ijk},
\label{10}\end{equation}
where $ \epsilon _{ijk}$ denotes the Levi-Civita antisymmetric
 tensor in Cartesian coordinates $( \epsilon _{123}= 1)$ \cite{2.}.
This geometry prepares the ground for the
 action principle to be developed.
\newpage
\noindent
{\bf 3)~Action
   principle
   for the rotational motion in the body-fixed $\mbox{system B}$}\\
Pure rotations around
 an arbitrary fixed point of the body are governed
 by  the Euler equations  (\ref{1}).
The kinetic energy is $T = \frac{1}{2} I_i  \Omega ^i  \Omega ^i$.
In S, the  Lagrangian is $L_S = \frac{1}{2}
   g^{\rm kin}_{ \mu  \nu } \dot \varphi ^ \mu
\dot  \varphi ^\nu - U( \varphi )$
 with $ U$ being the potential energy and $g^{\rm kin}_{ \mu  \nu }
   = I_i
 e^i{}_ \mu  e^i{}_\nu$ the ``kinetic-energy-metric". The
 latter metric differs
 from geometric metric
 $g_{ \mu  \nu }$ of Eq.~(\ref{x8}).\footnote{
Note that Einstein's summation
convention is extended to include also
diagonal components of the inertia tensor.}
 In this note, the
  kinetic-energy-metric $g^{\rm kin}_{ \mu  \nu }$
 \cite{8.} will play no role.
 The potential term $U (\varphi)$ is irrelevant for our considerations
 and will be dropped.

An application of Hamilton's action principle $ \delta A_S = 0$ to
 the classical action $A_S = \int^{t_2}_{t_1} L_S
 dt$  in the system $S$
certainly yields
the correct equations
of motion. Under the transformation  (\ref{4}),
they become the Euler equations (\ref{1}).

Let us now transform the action
$A_S$ via (\ref{4}) to the body-fixed system B.
The result is very simple:
\begin{equation}
A_B =
 \int^{t_2}_{t_1} dt L_B =
\int^{t_2}_{t_1} dt \frac{1}{2} I_i \dot \Phi ^i
       \dot \Phi ^i .
\label{11}\end{equation}
By applying naively
 Hamilton's action principle we would find the equations
$I_i \ddot \Phi ^i = I_i \dot \Omega ^i = 0$ for each
$i$ which are {\em not\/} the correct Euler
equations  (\ref{1}) --- the gyroscopic moments being missed.
The contradiction is caused by the fact that the variations $ \delta
\varphi ^\mu (t)$
in the space of Euler angles ${\rm M}_ \varphi $ and the variations
$\delta  \Phi ^i(t)$  in the space of anholonomic coordinates
${\rm M}_ \Phi $,
are related with each other
 in a
path dependent
way (``nonlocal" on the time axis) \cite{9.}.
Explicitly, there is the functional equation
\begin{equation}
   \Phi ^i_2 =  \Phi ^i_1 + \int^{t_2}_{t_1} e^i{}_ \mu
		  [ \varphi (t)] \dot  \varphi ^\mu (t) dt.
\label{12}\end{equation}
Its most important consequence is that
 closed paths in the space ${\rm M}_ \varphi $ are not
 mapped into  closed paths in the space ${\rm M}_ \Phi $,
  due to a nonvanishing Burgers vector $b^i = \oint e^i{}_\mu
  d \varphi ^\mu \neq 0$.

The usual Hamilton action principle in the system
$S$ proceeds by considering a variation $ \delta  \varphi ^\mu(t)
 = \bar \varphi ^\mu (t) -  \varphi^\mu (t)$ between two paths
$\bar  \varphi ^\mu (t),~ \varphi ^\mu (t)$   $(t \in [t_1, t_2 ])$
    with common  ends: $\bar  \varphi ^\mu (t_{1,2}) =  \varphi ^\mu
   (t_{1,2})$. Together, they form a {\em closed\/}
   path in the space ${\rm M}_ \varphi $.
The above naive
  application of Hamilton's action principle in the system B
employed analogous
closed-path
 variations $ \delta  \Phi ^i(t) =
\bar  \Phi ^i (t) -  \Phi ^i (t)$ between two paths
$\bar    \Phi ^i (t),  \Phi ^i (t)$  $(t \in [t_1, t_2])$,
with also  common ends: $\bar  \Phi ^i (t_{1,2}) =  \Phi ^i
(t_{1,2})$. This, however, cannot be correct. Under
a transformation  (\ref{4}), variations $ \delta
 \varphi ^\mu$ do  {\em not\/}
 produce closed-path variations $ \delta  \Phi ^i$ in the
space ${\rm M}_ \Phi $.
The nonzero Burgers vector $ { b}^i\neq 0$ of the anholonomy causes
a closure failure. This has to be accounted for
in a correct
derivation of the equations of motion.

The anholonomy of the transformation (\ref{4})
 requires distinguishing two types of variations of the paths
 in the space ${\rm M }_ \Phi$:
closed path
variations $ \delta  \Phi ^i (t)$ of the tpye described above, and
{\em anholonomic\/} variations $\deltabar  \Phi ^i (t)$,
 which are images of the $ \delta  \varphi ^\mu (t)$
variations in the space ${\rm M}_ \varphi $. Only the latter will
 produce
the correct equations of motion from an action principle
in the system B.

To calculate the variation  of the classical
action in the system B we derive the following simple
formula for the  anholonomic variation $ \deltabar \dot \Phi ^i$ of
the angular velocity $ \Omega ^i = \dot  \Phi   ^i$:
\begin{equation}
 \deltabar \dot \Phi ^i = \frac{d}{dt}
			    \delta \Phi^i
			  + \epsilon^i{}_{jk} \dot  \Phi^j
			     \delta \Phi^k
\label{13}\end{equation}
(corresponding to formula (9) in Ref.\ \cite{9.}).

Indeed, the transformation (\ref{4}) and the definition
 (\ref{5}) lead to the equation
\begin{equation}
 \deltabar  \dot \Phi  ^i =  \delta (e^i{}_ \mu  \dot \varphi ^ \mu )
     = e^i{}_ \mu   \delta  \dot \varphi ^ \mu  +  \partial _ \nu
	e^i{}_ \mu \dot \varphi ^ \mu   \delta  \varphi ^ \nu  =
	\frac{d}{dt} \left(e^i{}_ \mu  \delta  \varphi ^ \mu  \right)
	 + \left( \partial _ \mu e^i{}_ \nu -  \partial _ \nu
          e^i{}_ \mu  \right) \dot \varphi ^ \nu   \delta
\varphi ^ \mu
\label{14a}\end{equation}
which by (\ref{10a}) becomes
 \begin{equation}
\deltabar\Phi^i = \frac{d}{dt}   \delta \Phi^i
	+ 2 S_{kj}{}^{i} \dot\Phi^j   \delta \Phi^k
\label{}\end{equation}
leading to
(\ref{13}) via (\ref{10}).
The variations $ \delta \varphi^ \mu $ are
still performed
  in the  system S. They can be mapped {\em locally\/} into
  holonomic closed-path variations $ \delta  \Phi^i$ in
the system B by
  $ \delta \Phi^i = e^i{}_ \mu   \delta \varphi^ \mu $.
   They have the fixed-endpoint property $ \delta \Phi^i (t_{1,2}) = 0$,
  reflecting the closed-path condition
 $ \delta
  \varphi^ \mu (t_{1,2}) = 0$
 in the system S.

Let us emphasize that our final
  nonholonomic variations (15)
 are completely intrinsic to the system B.
Applying these
to the action $A_B$,
an  integration by parts
 with the boundary  conditions $ \delta  \Phi ^i  (t_{1,2})
   = 0$ leads to
\begin{equation}
\deltabar A_B  = \int^{t_2}_{t_1} dt I_i
   \dot\Phi^i \deltabar \Phi^i = - \int^{t_2}_{t_1}
   dt  \delta \Phi^i (\dot{\bf L} + {\bf  \Omega }
   \times {\bf L})^i
\label{}\end{equation}
Since $ \delta \Phi^i $
  are now ordinary (holonomic) variations,
 we find the Euler equations (1).
%
%
Thus  the anholonomic action principle
\begin{equation}
 \deltabar A_B = 0
\label{15}\end{equation}
 produces the correct
equations of motion
 in the
 body-fixed system B.
\\[5mm]
{\bf 4)~Action principle for general motion
   in the body-fixed $\mbox{system B}$}\\
To extend the action principle
to the  general motion of a rigid
  body one usually
chooses the point O in B to be the  center of mass  of the body and
studies the movements of O  and  rotations around O. Then
$I_j$ are the body's principal  moments of inertia.
The configuration space is a 6-dimensional manifold
 ${\rm M}^{(6)} = R^{(3)} \times {\rm SO}(3)$ and the kinetic
 energy
 $T = T_{\rm trans} + T_{\rm rot}$ consists of two terms:
 The first is the translational kinetic
energy $T_{\rm trans} = \frac{1}{2}M V^i V^i$ in the system B
the second is
   the rotational kinetic
  energy
$T_{\rm rot} = \frac{1}{2} I_i  \Omega ^i \Omega ^i$
in the system B.

In the system $S$, the position and  orientation of the body
are parametrized by the (holonomic) coordinates
$ \left\{  x ^ \mu , \varphi^ \mu \right\}_{ \mu  = 1,2,3}
  = \left\{ q^M \right\}_{M = 1,2, \dots 6}$,
with $ x ^ \mu $ being the mass  center
 coordinates and $\varphi^ \mu $ the Euler angles.

The body-fixed
basis vectors in B (see Fig.~1) are some function of the
Euler angles $\varepsilonb_i ( \varphi )$.
Their components in the cartesian basis
of the system $S$
are $\varepsilon_i{}^\mu (\varphi)$.
They form a $3 \times 3$ orthogonal matrix
${\varepsilon}  (\varphi) = ({\varepsilon}_i{}^ \mu  ),~
{\varepsilon}^{-1} =({\varepsilon}{}^i{}_ \mu) = {\varepsilon}^T$.
The components of the velocity of the center of mass
in the system $S$ are $ v^ \mu = \dot x^ \mu $. The
 components of $\dot x^ \mu $ with respect to the basis $\varepsilonb_i$
   are $V^i = {\varepsilon}^i{}_ \mu  \dot x^ \mu $.

The last relation  permits us to introduce a new
set of anholonomic coordinates $X^i$  describing
 the center of mass motion in the system S as seen from the system B.
By analogy with formula  (\ref{5}), we define the infinitesimal
increments $\dbar X^i$ as:
\begin{equation}
    \dbar X^i = \varepsilon^i{}_\mu (\varphi) dx^ \mu .
\label{16}\end{equation}
The set $\left\{  X^i,  \Phi ^i \right\}_{i =1,2,3}
    = \left\{ Q^I \right\}_{I  = 1, \dots , 6}$ gives a complete
set of an anholonomic coordinates for the configuration
space ${\rm R}^{(3)} \times {\rm SO}(3)$ of the moving  body.

Let us write down the anholonomic geometry of this space.
Introducing   $6 \times 6$ matrices,
whose rows and columns are labeled by capital Latin and Greek letters,
\begin{equation}
{\cal E} = \left(
\begin{array}{cc}
 e &  0\\
   0 &  \varepsilon
\end{array}
\right)
 =
 \left( {\cal E}^I{}_ \Lambda  \right),~~ {\cal G} =
 \left( {\cal G}_{ \Lambda  \Sigma }\right)
   =   {\cal E}^T {\cal E} = \left(
\begin{array}{cc}
  g  & 0\\
  0  & \hat 1
\end{array}
\right) , \dots
\label{}\end{equation}
the geometry possesses
the affine connection $  \Gamma _{ \Lambda \Sigma}{}^ \Delta  =
   {\cal E}_I{}^ \Delta   \partial  _ \Lambda {\cal E}^I{}_ \Sigma $
with the torsion tensor
$
S_{ \Lambda  \Sigma }{}^ \Delta  =  \Gamma _{[ \Lambda  \Sigma ]}{}^ \Delta
\neq 0$
and a vanishing Cartan curvature tensor
 $R_{ \Lambda  \Sigma , \Delta }{}^\Xi = 0$, implying
an nonvanishing Riemann curvature tensor $
 \bar R_{ \Lambda  \Sigma , \Delta }{}^\Xi \neq 0$.
This geometry is a simple extension of the 3-dimensional anholonomic
geometry on ${\rm SO}(3)$ described in Section 2. It reflects the structure
 of the present configuration space
 $R^{(3)} \times {\rm SO}(3) $, i.e., the
 space of configurations of the rigid body
    comprising translations of the
center of mass and rotations around it.

As a direct consequence of the definition (\ref{16}),
we have the relation $V^i =  \varepsilon^i{}_ \mu  \dot  x ^ \mu  =
 \dot X^i$  which, together with  (\ref{5}),
leads to the following simple form of the rigid body's
Lagrangian in the system B:

\begin{equation}
 L_B = \frac{M}{2} \dot X^i \dot X^i + \frac{1}{2}
   I_i \dot \Phi ^i \dot  \Phi ^i.
\label{17}\end{equation}

As before, a naive application of
 Hamiltons principle in the system B would produce a wrong
equation $M \ddot X^i = 0$ for the motion of the center-of-mass.
The correct anholonomic variation of
the velocity
\begin{equation}
 \deltabar \dot X^i = \frac{d}{dt} ( \delta X^i)
	 + \varepsilon^i{}_{jk}  \left( \delta \Phi^j \dot X ^k
           - \dot\Phi^j  \delta X^k \right).
\label{22}\end{equation}
Indeed, the transformation  (\ref{4a})
and the definition (\ref{16})  lead to
\begin{equation}
  \deltabar \dot X^i =  \delta (\varepsilon^i{}_ \mu
      \dot x^ \mu ) = \frac{d}{dt} ( \epsilon ^i{}_ \mu
        \delta x^ \mu ) + \partial _ \lambda  \epsilon ^i{}_ \mu
	  (\dot x^ \mu  \delta \varphi^ \lambda - \dot\varphi^ \lambda
	      \delta x^ \mu ).
\label{23}\end{equation}
The variations $ \delta x^ \mu $ may be mapped {\em locally\/}
  into holonomic closed-path variations $ \delta X^i
  =  \varepsilon ^i{}_ \mu  \delta x^ \mu $ in
  the system B  with the property $ \delta X^i(t_{1,2}) = 0$
  reflecting the closed path condition $ \delta x^ \mu (t_{1,2}) = 0$.
Formula (23) follows using
 the equations
 $\dot x^ \mu = \varepsilon^ \mu {}_k \dot X^k,~~  \delta  x^ \mu  =
\varepsilon^ \mu {}_k~ \delta X^k~~~ (\varepsilon^ \mu {}_k
\varepsilon^k{}_\nu =  \delta ^ \mu _ \nu ),~
 \dot\varphi^ \lambda  = e^ \lambda{}_j
 \dot\Phi^j,~~ \delta \varphi^ \lambda  = e^ \lambda {}_j
 \delta \Phi^j~~~ (e^ \lambda {}_j e^j{}_ \nu  =  \delta ^ \lambda _ \nu)_j$,
and the relation $(d\varepsilon  \varepsilon^T)_{ij} =
  -\varepsilon_{ijk} d\Phi^k$.

By applying the anholonomic variation (23)
to the translational energy, we obtain
%
%
\begin{equation}
\deltabar T_{\rm trans} =  \frac{d}{dt} (M \dot X^i
		     \delta X^i) - M(\ddot X^i + \varepsilon^i{}_{jk}
	     \dot\Phi^j \dot X^k)  \delta X^i.
\label{19}\end{equation}
An integration by parts  using of the fixed-ends
 conditions $ \delta  X ^ \mu (t_{1,2}) = 0$
leads
to the following expression for the total
anholonomic variation of the action   $A_B = \int^{t_2}_{t_1} L_B
dt$:
\begin{eqnarray}
   \deltabar A_B  & = & \int^{t_2}_{t_1} dt (
	I_i \dot\Phi^i \deltabar \dot\Phi + M \dot X^i
	  \deltabar \dot X^i) = \\
	& = & - \int^{t_2}_{t_1} dt [ \delta \Phi^i
		({\bf L}  + {\bf  \Omega } \times
		{\bf L})^i +  \delta X^i (\dot{\bf P}
		  + {\bf  \Omega } \times {\bf P})^i]
\label{20}\end{eqnarray}
From the second term, we find the equations (2)
for the translational motion.
%
%
\\[5mm]
5)~{\bf Conclusion}\\
By subjecting the action $A_B$ in the body-fixed system to
the new anholonomic variations with respect to
 translational
and rotational
 degrees of freedom
of the rigid body
we have been
 able to derive  both
the correct
 Euler
equations (\ref{1}) and the  equations (\ref{2})
completely within the body-fixed  system
without reference to the stationary
systems.

The existence of such an action principle
intrinsic to the body-fixed system
 may  not  be only of aesthetic  value but may also
 have  important practical consequences.
~\\                          ~\\~\\
Acknowledgement:
\\
We are greateful to Dr. S.V. Shabanov for useful discussions.

\vfill
\end{document}